\documentstyle[graphicx,float]{article}

\begin{document}
\title{Excluded states in entangled systems: technical and conceptual aspects}
\date{}
\author{Pedro Sancho \\ Centro de L\'aseres Pulsados CLPU \\ Parque Cient\'{\i}fico, 37085 Villamayor, Salamanca, Spain}
\maketitle
\begin{abstract}
The impossibility of ascribing definite states to the constituents
of an entangled system restricts the scope of Pauli's principle in
this context. We analyze the conceptual and physical aspects of the
problem by studying the actual scope of the principle and the
possibility of extending the concept of exclusion in correlated
systems. When the entanglement is weak, as in the archetypical case
of the Helium atom, the principle can be applied in an approximated
way. The concept of non-complete set of properties plays a crucial
role in the argument, which also clarifies the physical meaning of
the atomic quantum numbers; they are a multi-particle property, not
an one-electron one. In contrast, for strong entanglement the
excluded states are independent of the principle. We describe some
of these states, many times determined by symmetries of the
multi-fermion state.
\end{abstract}

\section{Introduction}

Pauli's exclusion principle \cite{Pau} is a fundamental tool to
describe the behavior of identical fermions. It states that two
identical fermions cannot be in the same physical state. Implicit to
this formulation is the assumption that individual fermions are in
definite states. This is a natural assumption for product states,
but becomes problematic for entangled ones. When a multi-particle
system is in an entangled state there is a state of the complete
system but there are not individual states of the particles
composing it. This well-known fact can also be formulated in terms
of properties of the particles: when the system is entangled one
cannot attribute complete sets of properties to the constituents
\cite{Gh1,Gh2,Gh3,ale}. If the constituents have not individual
states the exclusion conditions cannot be used in the standard way.

Technical and conceptual questions immediately arise from the above
considerations. We shall discuss both simultaneously in an
interconnected way. It is well-known that the principle has been
successfully applied to explain the structure of some entangled
systems as atoms. We shall show, in the case of the Helium atom,
that this fact can be explained because a weak entanglement
approximation works well. Within the weak entanglement approximation
it is possible, in some cases, to ascribe individual states to the
two electrons. Some of them are excluded by Pauli's principle,
explaining the energy levels structure of the atom. In addition, our
approach clarifies a subtle aspects of the problem. In the weak
entanglement regime it is possible to define non-complete sets of
spatial and spin properties. Although these partial properties
cannot define one-electron states, they can be used to justify the
definition of quantum numbers in the system. However, contrarily to
a naive expectation, these numbers are related to the full
two-electron state, not individually to each electron. The quantum
numbers are multi-particle properties.

After clarifying the above point another question appears in a
natural way: are there excluded states when the weak entanglement
approximation cannot be invoked? In such a regime of strong
entanglement Pauli's principle does not make sense. We answer this
question in the affirmative. One can define excluded states in the
same mathematical way  associated with Pauli-type ones: for some
conditions (equality of states in the standard case) the state
becomes undefined. We identify some of these generalized excluded
states, whose exclusion conditions are mainly related to the symmetries of
the multi-fermion state previous to antisymmetrization.

\section{States and properties in entangled systems}

Pauli's principle is a physical statement based on the states of the
constituents of the system. When the system is entangled the
definition of these states becomes a subtle subject, specially when
the components are identical particles. In this section we review
the main aspects of the problem.

Let us begin with the simpler case of distinguishable
(non-identical) particles. By the sake of simplicity we restrict our
considerations to two-particle systems. The extension to the general
case is straightforward. We consider a couple of particles with only
one degree of freedom, the spatial one (for instance, spin-$0$
particles). If we label $A$ and $B$ the two types of particles a
typical example of entangled state is
\begin{equation}
|\Psi >=\frac{1}{\sqrt 2}(|L>_A|R>_B - |R>_A|L>_B)
\label{eq:uno}
\end{equation}
with $L$ and $R$ denoting two non-overlapping, and consequently
orthogonal, spatial locations. From $\Psi$ we cannot ascribe
one-particle states to $A$ and $B$. We cannot say anything about the
state of $A$ without reference to the state of $B$, and vice versa.
The same conclusion can be reached via the reduced density matrix of
$A$, which is obtained by tracing out over $B$ the two-particle
density matrix $|\Psi><\Psi|$: $\rho _A =(|L><L|+|R><R|)/2$. This is
an improper mixture that cannot be associated with a pure state.
There are not one-particle pure states in entangled systems.

An alternative, and very powerful from the conceptual point of view,
way to analyze the subject was pioneered by Ghirardi et. al.
\cite{Gh1,Gh2,Gh3}. It is based on the consideration of properties
instead of states. In the case of systems in product states these
properties can be easily ascribed. For instance, in the state
$|L>_A|R>_B$ we can say that the particle $A$ ($B$) has the physical
property of being located at $L$ ($R$). A position measurement would
corroborate this assignment. In contrast, for the entangled state
$\Psi$ the localization of the particles is undetermined; they do
not have the physical property of being located at specific regions
of space. This approach is in the spirit of the
Einstein-Podolsky-Rosen argument, because measurements carried out
on the observable associated with the assigned property will give
with certainty that value.

The above conclusions, the absence of one-particle states and
properties for entangled systems, cannot be directly translated to
the case of identical particles. In this case the notion of
entanglement is much more subtle and there is no complete agreement
on its correct characterization yet. The reason for these
difficulties lies on the introduction of labels in the
symmetrization process that are not physical degrees of freedom. For
instance, a state of the type (\ref{eq:uno}) specialized to
identical particle reads $|\Psi _{ide}>=(|L>_1|R>_2 -
|R>_1|L>_2)/\sqrt 2$, where the labels $A$ and $B$ of
distinguishable particles have been replaced by $1$ and $2$,
indicating that now we are dealing with identical particles.
Although apparently the states $\Psi $ and $\Psi _{ide}$ share the
same type of  non-factorizability, there are actually fundamental
differences between them. The states like $\Psi _{ide}$ that
represent the (anti)symmetrization of a product state (in our
example $|L>_1|R>_2$) are not really entangled. It is easy to
demonstrate that they are not a resource of non-classical
correlations, the trademark of entangled states. This point has been
discussed in detail in \cite{Gh3,shp} for violations of Bell's
inequalities and in \cite{shp} for quantum information processing.
Thus, as signaled in \cite{shp}, the apparent entanglement of states
like $\Psi _{ide}$, suggested by their non-factorizability, is only
an artifact of a formalism with surplus structure. Nevertheless, it
must be stressed that this surplus character only refers to
entanglement. There are other physical processes, like interference,
that are sensitive to the (anti)symmetrized form of the states of
identical particles (see, for instance, \cite{shp,ps}).

All these and other considerations led to the definition of
entanglement for systems of identical particles proposed in
\cite{Gh1}: a system of identical particles is entangled when it is
impossible to ascribe a complete set of properties to each
constituent. Note that the set of properties must be complete, that
is, we must have a property for each degree of freedom of the
particle. This is the definition of entanglement of identical
particles we shall follow in the paper. Now, we are in position to
give an answer to the question of the existence of individual states
for entangled identical particles. As the physical properties of the
constituents are mathematically represented by one-particle states
this definition of entanglement automatically precludes the
existence of individual states.

If there are not states of the particles we will not be able of
individuating the constituents of the system. There have been many
discussions in the literature about how individuating quantum
particles \cite{die}. In classical physics we are able to
individuate even identical particles (with all their intrinsic
properties equal) via their positions. Unfortunately, this approach
fails in quantum theory, where localization is not so sharp. The
only possibility of individuating identical particles in the quantum
realm is via the relevant observables and pure states. The operators
associated with the observable and the pure states provide a
complete set of properties \cite{die}. When the system is entangled
these states simply do not exist, precluding any attempt of
distinguishing the constituents by physical means. The labels
associated with the (anti)symmetrization procedure are only
mathematical artifacts deserved of any connection with physical
properties.

The above discussion clearly indicates that we must rethink the
subject. The theory of excluded multi-fermion states, usually based
on one-fermion states and Pauli's principle, should be formulated in
a different way. As we shall discuss at length in the rest of the
paper such a formulation can be done based on two properties of the
system. The first one is the possibility of defining partial
properties of the constituents. The second one refers to exclusion
conditions based on properties of the multi-fermion state.

\section{Pauli's principle in the Helium atom}

The aim of this section is to justify the existence of quantum
numbers and the application of Pauli's principle in some entangled
atomic systems. Being these systems entangled, without individual
states of the electrons, the validity of these two properties is not
obvious at all. The analysis will show the importance of the concept
of partial sets of properties in the problem.

We restrict our considerations to the simplest case, the Helium
atom. The generalization to other atoms is simple. As it is
well-known, the states describing the two electrons in the Helium
atom are the product of the spatial and spin states, and they can be
written in three different forms
\begin{equation}
|\Psi _{\pm}>= N_{\pm}(\psi ({\bf x},{\bf y}) \pm \psi ({\bf y},{\bf
x}))(|s>_1 |s'>_2 \mp   | s'>_1 |s>_2)
\end{equation}
and
\begin{equation}
|\Psi _*>= N_*(\psi ({\bf x},{\bf y}) - \psi ({\bf y},{\bf x}))|s>_1
|s>_2
\end{equation}
with
\begin{equation}
N_{\pm}=(4 (1\pm Re(<\psi ({\bf x},{\bf y})|\psi ({\bf y},{\bf
x})>))(1\mp |<s| s'>|^2))^{-1/2}
\end{equation}
and
\begin{equation}
N_*=(2 (1 - Re(<\psi ({\bf x},{\bf y})|\psi ({\bf y},{\bf x})>)))^{-1/2}
\end{equation}
the normalization factors (provided that $\psi$ is normalized). For
$\Psi _+$ we symmetrize the spatial part and antisymmetrize the spin
one (and vice versa for $\Psi _-$ and $\Psi _*$). In all the cases
the complete state is antisymmetrized. At variance with textbook
presentations of the problem the spatial wavefunction has not a
product form, reflecting the mutual interaction of the electrons. In
the atomic physics jargon, $\Psi _-$ and $\Psi _*$ are ortho-Helium
type states and $\Psi _+$ is a para-Helium type one.

Note that we have used the antisymmetrization principle in its standard
form. The arguments supporting this principle are independent of the
factorizable or non-factorizable nature of the state. Then they are
valid for both entangled and product states.

The non-factorizable states of identical particles have two types of
contributions to the non-separability, those due to
antisymmetrization and to entanglement. It is necessary to
distinguish between them in order to correctly identify the true
entanglement. According to the criterion in \cite{Gh1} a two-fermion
state is non-entangled if and only if it is obtained by
antisymmetrizing a factorized state. Applying the criterion to the
previous examples we have that the states $\Psi _{\pm}$ are
entangled. On the other hand, $\Psi _*$ corresponds to the
antisymmetrization of the  state $\psi ({\bf x},{\bf y})|s>_1|s>_2$,
which is factorizable if and only if the wave function is a product
one, $\psi ({\bf x},{\bf y})=\psi ({\bf x})\phi ({\bf y})$. When
$\psi ({\bf x},{\bf y})$ is non-factorizable the state $\Psi _*$ is
entangled.

In general, the three above states are entangled preventing the use
of Pauli's principle in its usual form because of the absence of
one-electron states. However, a weak entanglement approximation
holds in the problem. When this approximation is valid we can define
partial spatial and spin properties of the electrons although the
one-electron states remain undefined. In the atomic case, these
partial properties are equivalent to the existence of quantum
numbers. Moreover, in that approximation the states $\Psi _*$ become
unentangled ones and we can use Pauli's principle. Next, we develop
this argument in detail.

First of all, we introduce the definition of the weak entanglement
regime. The usual description of the Helium atom considers in a
first approximation the two electrons without mutual interaction.
The states representing this first stage are separable, they are
equivalent to the states $\Psi $ but with the spatial part
factorized: $\psi ({\bf x},{\bf y})=\psi ({\bf x}) \phi  ({\bf y})$.
Later, the neglected mutual interaction is taken into account as a
perturbation, giving accurate energy levels. The ability of the
method to correctly reproduce the atomic structure shows the
validity of the approach and, in particular, of the use of product
spatial wave functions in these cases. The spatial entanglement must
be weak in order to write the wave function in a product form. Then
we can denote this method as the weak entanglement approximation.

When the approximation holds the $\Psi$'s can be expressed in the
(unnormalized) form $|\tilde{\Psi }_{\pm}> \sim (\psi ({\bf x})\phi
({\bf y}) \pm \psi ({\bf y}) \phi ({\bf x}))(|s>_1 |s'>_2 \mp   |
s'>_1 |s>_2)$ and $|\tilde{\Psi } _*> \sim (\psi ({\bf x}) \phi
({\bf y}) - \psi ({\bf y})\phi ({\bf x}))|s>_1 |s>_2$. According to
the criterion of \cite{Gh1} the spatial and spin parts of these
states are not entangled \cite{not} (although the full states
$\tilde{\Psi}_{\pm}$ are entangled). Then, following the
argumentation in \cite{Gh1}, we can attribute spatial and spin
properties to the parts of the system by separate. We can say that
one of the particles (we do not know which one) possesses the
spatial properties associated with the one-electron wave function $\psi$,
and the other those of $\phi$. Similarly, with respect to the spin
variables, we can say that one electron possesses the spin
projection $s$, and the other the $s$ projection (for $\tilde{\Psi
}_*$) or the $s'$ one (for $\tilde{\Psi }_{\pm}$). We remark that
for the states $\tilde{\Psi}_{\pm}$ we can define properties for the
spatial and spin variables but we cannot define a complete set of
properties for each one of the particles. For instance, for
$\tilde{\Psi}_-$, we have simultaneously the options for one of the
particles of possessing the properties of state $\psi$, with spin
projections $s$ or $s'$. We cannot attribute simultaneously spatial
and spin properties to the particle. The attribution of properties
is to the spatial and spin parts of the system or, equivalently, to
the two-particle system. This attribution of properties is objective
in the sense used in this paper \cite{no1}.

The possibility of attributing to the system the properties
associated with spatial wave functions implies that we can define
the usual quantum numbers ($n,l,m_l$). Similarly, for the spin
variable it is justified the definition of a spin projection (the
other usual quantum number $m_s$). It must be stressed that we do
not have a complete set of quantum numbers for each particle. One of
the electrons in $\tilde{\Psi}_-$ (we do not know which one) is
spatially characterized by $n,l,m_l$ and the other by
$n',l',m_{l'}$. In the same way, one of them possesses the property
$m_s$ and the other the $m_{s'}$ one. However, we cannot say if the
particle with the numbers $n,l,m_l$ is in the spin projection $m_s$
or in the $m_{s'}$ one. We only can assert that the two-particle
state possesses the set of quantum numbers
$n,l,m_l,n',l',m_{l'},m_s,m_{s'}$. The set of quantum numbers is a
property of the entangled two-electron state, not from the
individual electrons.

The situation is different for $\tilde{\Psi}_*$, where we can define
complete sets of properties for the particles. Each particle
possesses a defined set of quantum numbers and, moreover, a defined
state. This is a natural consequence of the fact that
$\tilde{\Psi}_*$ corresponds to the antisymmetrization of a product
state and is, consequently, non-entangled. Now, we can justify the
use of Pauli's principle in the Helium atom. It can be applied,
within the weak entanglement approximation, to all the states of the
form $\tilde{\Psi }_*$. It excludes these states when all the
quantum numbers are equal. The exclusion of all the states
$\tilde{\Psi}_*$ with equal quantum numbers is a necessary condition
to theoretically explain the observed level structure of the Helium.
The states $\tilde{\Psi}_*$ are only present in the structure of the
atom when some of the quantum numbers of $\phi$ differ from those of $\psi$,
for instance in the archetypical ortho-Helium state.

\section{Excluded states with strong entanglement}

When the weak entanglement approximation does not hold the Pauli
principle does not make sense and we must deal with the problem of
exclusion in a different way. Now, we are in the regime of strong
entanglement, where any reference to the states of the constituents
is misleading. In this section we show that the mathematical
definition of excluded states can be extended to the strong regime
and present some examples of these states.

The definition of excluded states (when Pauli's principle can be
applied) can be done in terms of their mathematical properties. Let
us consider the general two-fermion antisymmetrized state
\begin{equation}
\frac{|\psi>_1|\phi>_2-|\phi>_1|\psi>_2}{(2-2|<\psi |\phi
>|^2)^{1/2}}
\end{equation}
When $\psi = \phi$ the two-fermion state becomes an undetermined
expression of the type $0/0$. This is a quantitative expression of
Pauli's principle. From now on, we say that a multi-fermion state is
excluded when for some condition the state becomes in the
undetermined form $0/0$. In the case of Pauli's principle that
condition is the equality of the individual states.

We present next some examples of these excluded states. We consider
again the states $\Psi _\pm$ and $\Psi _*$, but now not necessarily
referring to the two electrons of the Helium atom. We start with the
state $\Psi _-$. When the spatial wave function is symmetric, $\psi
({\bf x},{\bf y})= \psi ({\bf y},{\bf x})$, the numerator and the
denominator are both null ($N_-=1/0$), becoming $\Psi _-$ undefined.
Clearly this is a manifestation of an exclusion-type behavior: a
condition on the state of the system precludes its preparation. The
difference with product states is that now the condition does not
refer to the states of the particles but to the state of the
complete two-particle system. The reasoning for $\Psi _*$ is
similar. When the spatial wave function is symmetric the state is
undefined.

The case $\Psi _+$ is different. Now, we have two possibilities. The
first one is for an antisymmetric spatial wave function, $\psi
({\bf x},{\bf y})= -\psi ({\bf y},{\bf x})$, when we obtain again an
undefined state. The second possibility corresponds to equal spin
projections, $s'=s$, which also implies an undefined state. Note
that in the two situations (just as for $\Psi _-$ and $\Psi _*$) the
exclusion conditions only refer to one of the two variables, spin or
spatial ones, whereas for the standard principle the conditions
always simultaneously refer to the two variables.

The generalization of excluded states is by no means restricted to
two-fermion systems. We briefly discuss the general case of $n$
identical fermions. We show that, at the level of sufficient exclusion
conditions, it is only necessary to consider two-fermion symmetries.
This property easily follows from a well-known textbook argument.
The state of the $n$ fermions previous to antisymmetrization is
denoted by $\chi $, from which we obtain via the usual
antisymmetrization procedure the (unnormalized) state $\varphi$. The
permutation of the $i$ and $j$ variables of this state leads to
\begin{eqnarray}
|\varphi ({\bf x}_1,s_1; \cdots ; {\bf x}_i,s_i; \cdots ; {\bf x}_j,s_j; \cdots ; {\bf x}_n,s_n)>= \nonumber \\
-|\varphi ({\bf x}_1,s_1; \cdots ; {\bf x}_j,s_j; \cdots ; {\bf
x}_i,s_i; \cdots ; {\bf x}_n,s_n)> \label{eq:ant}
\end{eqnarray}
The state $\chi $ is symmetric with respect to the fermions $i$ and $j$ if
\begin{eqnarray}
|\chi ({\bf x}_1,s_1; \cdots ; {\bf x}_i,s_i; \cdots ; {\bf x}_j,s_j, \cdots ; {\bf x}_n,s_n)> = \nonumber \\
|\chi ({\bf x}_1,s_1; \cdots ; {\bf x}_j,s_j ; \cdots ; {\bf
x}_i,s_i; \cdots ; {\bf x}_n,s_n)>
\end{eqnarray}
When this property holds for $\chi $ it is immediate to verify that
the antisymmetrized state vanishes, $|\varphi >=0$. Taking into
account that in the normalized form the state reads $|\varphi >/|
|\varphi >|=0/0$, we obtain again the undetermined expression
characterizing excluded states. The above exclusion condition refers
to a two-fermion symmetry of the complete multi-fermion state. This
condition translates in the case of product states to the usual
equality of the states of two particles: for a factorizable state,
$|\chi> =|\chi _1> \cdots |\chi _n>$, the $ij$-symmetry condition is
equivalent to the relation $|\chi _i> = |\chi _j>$. The equality of
states for separable systems can be seen as a particular case of the
more general symmetry relation.

The two-fermion symmetry relation is a sufficient exclusion
condition. Next we provide an example where it is not a necessary
one. We consider the case where the spatial and spin parts of the
state can be factored, $|\varphi >=\Upsilon ({\bf x}_1, \cdots ,
{\bf x}_n )|\upsilon (s_1, \cdots , s_n)>$, with $\Upsilon$
antisymmetric and $\upsilon$ symmetric. The spatial wave function
corresponds to the antisymmetrization of $\xi$. If $\xi$ is
symmetric with respect to the fermions $i$ and $j$, $\xi ({\bf x}_1,
\cdots , {\bf x}_i, \cdots , {\bf x}_j, \cdots , {\bf x}_n) = \xi
({\bf x}_1, \cdots , {\bf x}_j, \cdots , {\bf x}_i, \cdots , {\bf
x}_n)$, the state is forbidden. The $ij$-symmetry of the spatial
wave function, not of the complete state, is a sufficient exclusion
condition. Note also that once more, in contrast with the Pauli
principle, it is not necessary to introduce conditions on the spin
variables to exclude the state.

\section{Conclusions}

We have studied in this paper the meaning of exclusion in systems
composed of entangled identical fermions. The impossibility of
defining states for the constituents prevents the application of the
usual approach based on Pauli's principle. Only when the
entanglement is weak one can work, in an approximate way, in that
standard framework. In that approximation, although there are not
individual states, one can define partial properties for each one of
the variables involved in the problem. These partial properties are
equivalent in the atomic case to the introduction of quantum
numbers, justifying our argument its use in entangled systems.
However, these numbers cannot be understood in the standard way.
They are not the labels of one-electron properties. Instead, they
characterize the multi-electron state. They allow us to classify the
different atomic states according to these labels. The other
important consequence of the weak entanglement approximation is that
the states $\Psi _*$ can be represented by unentangled ones
$\tilde{\Psi }_*$, within the scope of Pauli's principle. The
exclusion of those with equal quantum numbers gives a consistent
description of the electronic structure of the Helium.

When we do not even have the possibility of defining partial
properties we must confront the problem from a different
perspective. In order to replace the original idea of exclusion, we
have explored a concept that can extend it. The extended
excluded states, are the natural mathematical generalization of the
standard principle. It refers to states that for some condition have
the undefined $0/0$ form typical of Pauli-type systems (when the
states of the components are equal). In the examples considered in
this paper the generalized exclusion conditions are mostly associated with
the symmetries of the multi-fermion state previous to antisymmetrization.

In this work we have not been only interested in the physical
aspects of the problem but also in the conceptual ones. Our analysis
has highlighted the advantages of studying the subject from the
point of view of the properties. This approach makes clear why there
are not individual states in entangled systems of identical
particles, a property that precludes the individuation of the
constituents. Nevertheless, the most subtle aspect of the conceptual
analysis is probably the status of the sets of partial properties;
we have seen that they can only be associated with the complete
system not with the constituents.

We conclude that the concept of exclusion must be viewed in a
different way in entangled systems. The present work is only a first
contact with the problem. Many other questions must be addressed. A
first obvious issue is to provide an exhaustive classification of
the physical conditions that lead to undetermined states of the type
$0/0$. Another line of research is to consider the possible
existence of other types of forbidden states, for instance, those
associated with consistency conditions related to the presence of
entanglement. We expect the unexplored regime of strong entanglement
to be a source of interesting novel aspects of exclusion.
Independently of any future development, the results presented here
show once more the importance of studying the interplay between
entanglement and particle identity.


\begin{thebibliography}{99}
\bibitem{Pau} W. Pauli, {\it Z. Physik} 1925, {\bf 31,} 765.
\bibitem{Gh1} G. C. Ghirardi, L. Marinatto, and T. Weber, {\it J. Stat. Phys.} 2002 {\bf 108,} 49.

\bibitem{Gh2} G. C. Ghirardi and L. Marinatto, {\it Fortschr. Phys.} 2003 {\bf 51,} 379.
\bibitem{Gh3} G. C. Ghirardi and L. Marinatto, {\it Phys. Rev. A.} 2004 {\bf 70,} 012109.
\bibitem{ale} M. C. Tichy, F. Mintert, and A. Buchlleitner, {\it J. Phys. B} 2011 {\bf 44,} 192001.
\bibitem{shp} J. Ladyman, ${\O}$. Linnebo, and T. Bigaj, {\it Studies in History and Philosophy of Modern Physics.} 2013 {\bf 44,} 215.
\bibitem{ps} P. Sancho, {\it Phys. Rev. A} 2010 {\bf 82,} 033814.
\bibitem{die} D. Dieks, Quantum Information and Locality {\it What is Quantum Information?} 2017 O. Lombardi, S. Fortin, F. Holik, and C. L\'{o}pez (editors), Cambridge University Press.
\bibitem{not} Note that we symmetrize one of the parts of the state (the spatial or the spin one) and antisymmetrize the other, being the full state antisymmetric. To determine if the symmetrized part is entangled or not we must use the criterion for bosons in \cite{Gh1}. Then the symmetrized part is separable if and only if it can be obtained by symmetrization of a factorized product of two orthogonal states or it is the product of the same state for the two particles. In our case this is so. For the  spin part of the state $\tilde{\Psi}_-$ because $s$ and $s'$ are orthogonal. Similarly, for $\tilde{\Psi}_*$, it is the product of the same state $s$. On the other hand, for the spatial part of the state $\tilde{\Psi}_+$, $\psi $ and $\phi$ are orthogonal when they correspond to different energy eigenstates. If not, they are the same state and we can apply the second part of the criterion.
\bibitem{no1} In order to make this point more intuitive we consider the
state $(|L>_1|R>_2-|R>_1|L>_2)(|s>_1|s'>_2+|s'>_1|s>_2)/2$. In the
spirit of \cite{Gh1} the ascription of spatial properties to the two-particle system is objective when it can be corroborated by measurements. In this
example simultaneous position measurements at $L$ and $R$ would
certify the presence of one particle at each side. A similar
argument holds for the spin variable.
\end{thebibliography}
\end{document}